\documentclass[12pt]{article}
\pdfoutput=1 

\usepackage{cancel,slashed}
\usepackage{bbm}
\usepackage{amssymb}
\usepackage{amsfonts}
\usepackage{amsmath}
\usepackage{graphicx}
\usepackage{cite}

\usepackage{latexsym}
\usepackage{color}
\usepackage{xypic}
\usepackage{cancel,slashed}
\usepackage[linktocpage]{hyperref}
\usepackage{color}
\usepackage{transparent}
\usepackage{footmisc}

\makeatletter
\def\@xfootnote[#1]{%
  \protected@xdef\@thefnmark{#1}%
  \@footnotemark\@footnotetext}
\makeatother

\newcommand{\bmat}{\left(\begin{array}}
\newcommand{\emat}{\end{array}\right)}

\def\p{\partial}
\def\a{\alpha}

\def\b{\beta}
\def\g{\gamma}
\def\d{\delta}

\def\-{\hphantom{-}}

\def\s2{\frac{1}{\sqrt2}}

\def\oh{\frac{1}{2}}
\def\beq{\begin{equation}}
\def\eeq{\end{equation}}
\def\beqa{\begin{eqnarray}}
\def\eeqa{\end{eqnarray}}

\def\tr{{\rm tr \,}}

\def\Dsl{\,\raise.15ex\hbox{/}\mkern-13.5mu D} 

\def\CM {{\cal M}}

\def\CL {{\cal L}}

\def\tr{\mbox{Tr}}

\def\be{\begin{equation}}
\def\ee{\end{equation}}
\def\bea{\begin{eqnarray}}
\def\eea{\end{eqnarray}}

\def\IZ{\mathbb{Z}}
\def\IR{\mathbb{R}}

\def\oh{\frac{1}{2}}
\def\a{{\alpha}}
\def\b{{\beta}}
\def\d{{\delta}}

\def\sig{{\sigma}}

\def\g{{\gamma}}

\def\p{{\partial}}

\def\w{{\wedge}}

\newcommand{\mr}{\mathrm}
\newcommand{\mc}{\mathcal}

\newcommand{\mb}{\mathbb}

\def\sm2{{\mbox{\small 2}}}

\newcommand{\bp}{\begin{pmatrix*}[r]}  
\newcommand{\ep}{\end{pmatrix*}}  
\newcommand{\bpp}{\begin{pmatrix}}  
\newcommand{\epp}{\end{pmatrix}}  
\newcommand{\bcd}{\begin{center}
\begin{tikzcd}}
\newcommand{\ecd}{\end{tikzcd} \end{center}}

\def\P{\mathbb{P}}

\def\C{\mathbb{C}}

\def\1{\mathbb{1}}

\def\h{{\rm h}}
\def\p{{\rm p}}
\def\d{{\rm d}}
\def\del{\partial}
\def\delbar{\bar{\partial}}

\def\L{\mathcal{L}}
\def\C{\mathcal{C}}
\def\M{\mathcal{M}}
\def\P{\mathcal{P}}

\begin{document}
\pagestyle{plain}

\makeatletter
\@addtoreset{equation}{section}
\makeatother
\renewcommand{\theequation}{\thesection.\arabic{equation}}
\pagestyle{empty}
\rightline{ IFT-UAM/CSIC-17-063}
\vspace{0.5cm}
\begin{center}
\Huge{{Compact T-branes}
\\[15mm]}
\normalsize{Fernando Marchesano,$^1$ Raffaele Savelli,$^{1,2}$ and Sebastian Schwieger$^1$ \\[10mm]}
\small{
${}^1$ Instituto de F\'{\i}sica Te\'orica UAM-CSIC, Cantoblanco, 28049 Madrid, Spain \\[2mm] 
${}^2$ Departamento de F\'{\i}sica Te\'orica, 
Universidad Aut\'onoma de Madrid, 
28049 Madrid, Spain
\\[8mm]} 
\small{\bf Abstract} \\[5mm]
\end{center}
\begin{center}
\begin{minipage}[h]{15.0cm} 

We analyse global aspects of 7-brane backgrounds with a non-commuting profile 
for their worldvolume scalars, also known as T-branes. In particular, we consider 
configurations with no poles and globally well-defined over a compact K\"ahler surface.
We find that such T-branes cannot be constructed on surfaces of positive or vanishing Ricci 
curvature. For the existing T-branes, we discuss their stability as we move in K\"ahler 
moduli space at large volume and provide examples of T-branes splitting into 
non-mutually-supersymmetric constituents as they cross a stability wall.

\end{minipage}
\end{center}
\newpage
\setcounter{page}{1}
\pagestyle{plain}
\renewcommand{\thefootnote}{\arabic{footnote}}
\setcounter{footnote}{0}


\tableofcontents


\section{Introduction}
\label{s:intro}

One important feature of type II string compactifications is the amount of information on the effective, lower-dimensional theory that one obtains by analysing BPS D-branes. In fact, knowledge on the spectrum of BPS D-branes is a requirement to build interesting type II string vacua, since they typically host the non-trivial gauge sector of the compactification \cite{thebook}. The more precise this knowledge is, the better the picture on the set of vacua on a certain region of the string landscape.

While D-brane BPS conditions have been thoroughly analysed for different classes of vacua, solving them explicitly can oftentimes be challenging. In that sense, a particularly tractable set of vacua is given by type IIB Calabi-Yau orientifolds with O3/O7-planes. Indeed, in this case the set of space-time-filling BPS D-branes at large volume is given by D3-branes and D7-branes. On the one hand, the embedding of a single D3-brane is simply a point in the internal six-manifold $B$, which trivially satisfies the BPS conditions. On the other hand, the BPS conditions for a single D7-brane demand that it wraps a holomorphic four-cycle $S \subset B$, threaded by an anti-self-dual worldvolume flux. Thanks to the machinery of K\"ahler geometry, finding the full set of such four-cycles and fluxes for a compact Calabi-Yau is a relatively easy task. A similar statement applies to stacks of D7-branes on $S$ endowed with non-Abelian, anti-self-dual gauge bundles, which allows to have a good grasp on the spectrum of space-time-filling BPS D-branes in this setting. Overall, the capability to construct such D-branes explicitly explains to great extent why type IIB orientifold compactifications and their F-theory generalisation has flourished so much in the past few years \cite{Denef:2008wq,Heckman:2010bq,Weigand:2010wm,Wijnholt:2012fx,Leontaris:2012mh,Maharana:2012tu,Quevedo:2014xia}. 

Nevertheless, it is precisely in this context where it has been realised that such BPS solutions may not be sufficient to realise certain phenomenological features. In addition to the above configurations, one may consider stacks of 7-branes whose complex worldvolume scalar (a.k.a. Higgs field) has a non-Abelian profile. These new objects are typically known as Higgs bundles in the mathematics and T-branes in the string theory literature \cite{Donagi:2003hh,Hayashi:2009bt,Cecotti:2010bp,Donagi:2011jy}, and can be seen as generalisations of the original construction of Hitchin \cite{Hitchin:1986vp}. As pointed out in \cite{Hayashi:2009bt,Cecotti:2010bp}, such T-brane backgrounds are crucial to engineer and compute realistic Yukawas in F-theory GUTs \cite{Donagi:2008ca,Beasley:2008dc,Beasley:2008kw,Donagi:2008kj}, as checked in explicit models in \cite{Chiou:2011js,Font:2013ida,Marchesano:2015dfa,Carta:2015eoh}. Since then there has been a lot of effort in understanding this class of backgrounds \cite{Donagi:2011dv,Anderson:2013rka,DelZotto:2014hpa,Collinucci:2014taa,Collinucci:2014qfa,Cicoli:2015ylx,Collinucci:2016hpz,Bena:2016oqr,Marchesano:2016cqg,Mekareeya:2016yal,Ashfaque:2017iog,Anderson:2017rpr,Bena:2017jhm,Collinucci:2017bwv,Cicoli:2017shd}.

Unlike their Abelian counterparts, configurations of 7-branes with a non-Abelian Higgs-field profile are poorly understood, in the sense that it is not known when they can be accommodated in type IIB/F-theory compactifications. Needless to say, this knowledge is necessary to realise the full model building potential of this class of vacua. This paper aims to make progress in this direction by analysing the conditions to construct T-branes with a compact embedding. That is, we analyse D7-branes with a non-Abelian profile for its worldvolume scalar $\Phi$, globally well-defined over a compact K\"ahler four-cycle $S$ and without any poles. We dub such configurations as {\it compact T-branes}, and analyse them by inspecting the related Hitchin system of equations over $S$. We therefore extend previous analysis of this sort, which so far have been essentially performed only at a local level.\footnote{An alternative treatment is via tachyon condensation techniques, particularly suitable for T-branes defined over 7-brane intersections. In this case a global analysis can also be carried out, as shown in \cite{Collinucci:2014qfa}.}

As usual, obstructions may be found when trying to extend a local solution globally. In our case we find that constructing compact T-brane solutions crucially depends on the Ricci curvature of the surface $S$, and more precisely on its cohomology class. Indeed, we find obstructions to the existence of compact T-branes over complex four-cycles of vanishing or positive-definite curvature, like K3 or del Pezzo surfaces. On surfaces of negative-definite curvature, instead, solutions can always be constructed, generalising the result of Hitchin for Riemann surfaces of genus $g >1$ \cite{Hitchin:1986vp}. Finally, for surfaces of indefinite curvature the construction will depend on the particular region of the K\"ahler moduli space where we sit.\footnote{More precisely, we find that, if $\rho$ is the Ricci form of $S$ and $J$ its K\"ahler form, then compact T-branes can be constructed when $\int_S \rho \wedge J < 0$.}  This latter case raises the question of the fate of T-branes when we move in K\"ahler moduli space and, in particular, when we pass from one region to another by crossing stability walls. In this respect, we find that a T-brane is either converted into a different BPS object as it crosses the wall, or it splits into non-mutually-BPS constituents. As could be expected, the T-brane's fate will ultimately depend on its topological data, and we analyse several interesting cases in terms of them. 

The paper is organised as follows. In section \ref{sec:compact} we specify the class of T-branes that we will be studying, with special emphasis on their global description in terms of a compact four-cycle. We then turn to discuss solutions to the BPS equations, first the analogous of the original Hitchin solution and then generalisations thereof. In section \ref{sec:nogo} we prove a topological obstruction to building compact T-brane solutions: they cannot be hosted by four-cycles of vanishing or positive-definite Ricci curvature class. Finally, in section \ref{sec:stability} we analyse the stability of the allowed T-brane constructions as we move in large volume K\"ahler moduli space, and in particular their fate after crossing a stability wall. We draw our conclusions in section \ref{sec:conclu}. 

Some technical details are relegated to the appendices. In appendix \ref{ap:alpha} we give a four-dimensional interpretation of the non-harmonicity of the worldvolume flux in T-brane solutions. In appendix \ref{ap:examples} we construct several explicit examples of the stability-wall transitions discussed in section \ref{sec:stability}.

\section{Global aspects of T-branes}
\label{sec:compact}

Consider a stack of 7-branes wrapping a compact K\"ahler surface $S$. Following \cite{Donagi:2008ca,Beasley:2008dc,Beasley:2008kw,Donagi:2008kj}, the 7-brane configuration and degrees of freedom can be characterised in terms of an eight-dimensional action on $\IR^{1,3} \times S$ with a non-Abelian symmetry group $G$. In particular, such data are encoded in terms of two two-forms on $S$: the field strength $\mathbb{F} = d\mathbb{A} - i\mathbb{A} \wedge \mathbb{A}$ of the 7-branes gauge boson $\mathbb{A}$, and the (2,0)-form Higgs field $\Phi$, whose eigenvalues describe the 7-brane transverse geometrical deformations. Both $\mathbb{A}$ and $\Phi$ transform in the adjoint of the initial gauge group $G$, which is nevertheless broken to a subgroup due to their non-trivial profile. Finally, such profiles need to satisfy certain equations of motion, which in the case of supersymmetric configurations are given by
\begin{subequations}
\label{susy7}
\begin{align}
\label{Fterm1}
\bar \partial_{\mathbb{A}} \Phi = &\, 0\\
\label{Fterm2}
\mathbb{F}^{(0,2)} =&\, 0\\
\label{Dterm}
J \wedge \mathbb{F} +\frac{1}{2} [\Phi, \Phi^\dagger] =&\,0\,,
\end{align}
\end{subequations}
where $J$ is the K\"ahler two-form of $S$. These equations are a generalisation of the celebrated Hitchin system \cite{Hitchin:1986vp} to a four-manifold. Upon dimensional reduction to four dimensions, the first two equations ensure the vanishing of the F-terms, while the third equation ensures the vanishing of the D-terms.

In this paper we will analyse 7-brane backgrounds with non-commuting expectation values for the worldvolume scalar $\Phi$, namely such that $[\Phi, \Phi^\dagger] \neq 0$, also known as T-branes in the string theory literature. We will restrict to those T-brane configurations that are globally well-defined over a compact K\"ahler surface $S$ and such that the Higgs field profile is absent of poles.\footnote{See \cite{Anderson:2017rpr} for a recent account of Hitchin systems with poles.}  We dub such T-brane configurations as {\em compact T-branes}, in the sense that the spectral equation for $\Phi$ describes a compact surface. Notice that poles are naturally associated to field-theory defects originating from additional 7-branes intersecting the stack, so we may interpret a compact T-brane as a stack of 7-branes in isolation from the others. In other words, we may see them as basic building blocks of BPS 7-brane configurations in type IIB/F-theory compactifications. We will moreover focus on solutions of equations \eqref{susy7} involving an Abelian profile for the gauge field. 
Said differently, in our backgrounds the source of non-commutativity of the 7-brane system will come entirely from $\Phi$.

In order to describe the essential features of compact T-branes, in this section we will focus on the simplest possible example, namely a stack of two D7-branes. This case allows to generalise the original example of Hitchin on a Riemann surface \cite{Hitchin:1986vp} to a compact complex four-cycle. From there one may generalise the T-brane Ansatz in a number of ways, finding backgrounds with a non-harmonic worldvolume flux. As we will see, the departure from harmonicity is governed by certain non-linear differential equations, and this will allow to connect our constructions with the literature of T-brane solutions in flat space. 

\subsection{T-branes and non-harmonic fluxes}\label{ss:global}

Let us focus on a stack of two 7-branes wrapping $S$, and therefore on a super-Yang-Mills theory on $\IR^{1,3} \times S$ with symmetry group $G = SU(2)$. We will always assume that $S$ is simply-connected, i.e. $\pi_1(S)=0$. This will simplify our analysis considerably because it implies, in particular, that holomorphic line bundles on $S$ have their topology completely specified by the first Chern class.  As mentioned, we will also restrict attention to a rank-two gauge bundle $\mathbb{V}$ on $S$ of split type, i.e.
\be
\mathbb{V}=\mathcal{L}\oplus\mathcal{L}^{-1}\,,
\label{SplitV}
\ee
where $\mathcal{L}$ is a line bundle whose curvature we denote by $F$. The F-term (\ref{Fterm2}) of the eight-dimensional super-Yang-Mills theory forces $F$ to be a differential form of Hodge-type $(1,1)$, which gives $\mathcal{L}$ a holomorphic structure. Moreover, since $F$ is closed, using the Hodge decomposition, we can uniquely write it as
\be\label{Alpha}
F=F^{\h}+\d\alpha\,,
\ee
where the superscript $^{\h}$ denotes the harmonic representative and $\alpha$ is a globally well-defined one-form. Note that the absence of non-trivial first-cohomology classes on $S$, following from its simply-connectedness, forbids harmonic representatives for $\alpha$. We can thus always choose (globally) a gauge that kills the exact part of $\alpha$, such that we can write
\be\label{generalalpha}
\alpha=-\frac{{\d}^c \,g \,({\bf x},{\bf \bar{x}})}{2}\,,
\ee 
where $g({\bf x},{\bf \bar{x}})$ is a globally well-defined real function on $S$ (with local complex coordinates collectively denoted by ${\bf x}$) such that $\int_S g\, \d {\rm vol}_S = 0$, and $\d^c=i(\delbar-\del)$. Using that $S$ is K\"ahler, it is easy to see that the co-differential operator $\delta=-*\d*$ annihilates the expression \eqref{generalalpha}, and hence $\alpha$ is co-closed. In this way, the gauge field strength becomes
\be\label{Fsplit}
F=F^{\h}-i\del\delbar g\,.
\ee

The function $g$, or equivalently $\alpha$, will play a key r\^ole in the sequel.  It will be the unknown of the non-linear partial differential equation governing T-brane backgrounds, which arises from the equation \eqref{Dterm} of the eight-dimensional super-Yang-Mills theory. In an ordinary intersecting-brane background, where $\Phi$ is diagonalisable, this equation forces $F$ to be primitive. By a standard result in K\"ahler geometry (see e.g. \cite{Ballmann}), every primitive (1,1)-form on a K\"ahler two-fold is anti-self-dual with respect to the Hodge-star operator. Since $F$ is closed, this implies then that $F$ is also co-closed, and hence harmonic. 
Now, reversing the argument, a T-brane supersymmetric configuration will involve a gauge field strength which is closed but not anti-self-dual, and therefore $F$ will not necessarily be given by the harmonic representative of a certain cohomology class. This departure from harmonicity is described by $g$.

As we will see, the information that $g$ encodes is lost in the four-dimensional effective theory. It can only be recovered when we include the D7-brane Kaluza-Klein modes into the four-dimensional description, as we discuss in appendix \ref{ap:alpha}. In other words, $g$ determines the microscopic details of the T-brane background, which only the eight-dimensional theory is sensitive to.

In order to determine  $g$ let us for convenience define the global real function
\be\label{varphi}
\varphi({\bf x},{\bf \bar{x}})\sigma_3\equiv *[\Phi,\Phi^\dagger]\,,
\ee
where, compatibly with our choice of gauge bundle $\mathbb{V}$, we restrict our attention to commutators proportional to the third Pauli matrix $\sigma_3$. Then one can see that $\varphi \geq 0$ all over $S$ and that equation \eqref{Dterm} reads
\be\label{8dDterm}
F\wedge J=- \frac{\varphi}{4} \,J^2\,.
\ee
Using the Lefschetz decomposition of harmonic forms, we can write
\be\label{SplitPrim}
F^{\h}=\frac{c}{4} \, J + F^{\h}_{\p}\,,
\ee 
where $c$ is a constant, $F^{\h}_{\p}$ is primitive and the numerical factor is for later convenience. Of course this splitting depends on the K\"ahler moduli of our string compactification, and the periods of the two summands are generally real (moduli-dependent) numbers which must add up to (half-)integer numbers to satisfy the quantization condition for $F$.\footnote{Recall that, in cohomology, $\tfrac{1}{2\pi}[F]=c_1(\mathcal{L})$.}

Using that $S$ is K\"ahler, one can show that $2i\del\delbar g\wedge J=*\Delta g$, where $\Delta$ is the Laplace operator in real coordinates. This leads us to an elegant rewriting of equation \eqref{8dDterm}:
\be\label{DiffD}
\Delta g({\bf x},{\bf \bar{x}}) = c + \varphi({\bf x},{\bf \bar{x}})\,.
\ee

At this point, one fixes an hermitian metric on $S$, and solves equation \eqref{DiffD} for $g$, or equivalently for the unitary connection $A$ on $\mathcal{L}$. Notice that a necessary requirement to solve this equation is that its r.h.s. integrates to zero, i.e.
\be\label{4dDterm}
c=-\frac{1}{{2\rm Vol}(S)}\,{\rm Tr}\int_S[\Phi,\Phi^\dagger]\,\sigma_3\,,
\ee
which is nothing but the condition for vanishing D-term potential in the four-dimensional low-energy effective theory.

Practically, equation \eqref{DiffD} can only be solved analytically in few situations, because in general $\varphi$ will depend non-linearly on $g$. Nevertheless this equation is always of elliptic type \cite{Hitchin:1986vp} and, as such, on a compact manifold it admits a unique smooth solution if the input function $\varphi$ is smooth and provided that \eqref{4dDterm} is satisfied \cite{Vafa:1994tf}.

The most convenient and adopted \cite{Cecotti:2010bp,Donagi:2011jy} approach to formulate the problem is to fix the holomorphic structure of $\L$ such that $A^{0,1} = 0$, which turns the anti-holomorphic covariant derivative of equation \eqref{Fterm1} into the simple Dolbeault operator $\delbar$. In this frame, equation \eqref{Dterm} (or else \eqref{DiffD}), becomes an equation for the hermitian metric $h$ on $\L$, which appears in the gauge field strength. The latter is indeed the curvature of the associated Chern connection $A^{1,0}\sim h^{-1}\del h$, i.e. locally $F=-i\del\delbar \log h$. Given that we can locally write $F^\h=-i\del\delbar \log h_0$ and that $F$ and $F^\h$ are in the same cohomology class, we see that the unknown function $g$ is globally-well defined and enters the metric $h$ as a conformal factor, i.e. $h=h_0\, e^g$.

For concreteness, let us consider a nilpotent Higgs field profile
\be\label{NilpPhi}
\Phi=\left(\begin{array}{cc}
0&m\\ 0 &0
\end{array}\right)
\ee
where $m \in H^{2,0}(S,\CL^{2})$. Equivalently, we can also see $m$ as a scalar holomorphic section of the line bundle $\CM \equiv \mathcal{L}^2\otimes K_S$, with $K_S$ the canonical bundle of $S$. By a slight abuse of notation, in the following we will describe both kinds of object with the same symbol, being clear from the context which one we are referring to. As it stands, this profile is a solution of equation (\ref{Fterm1}) in the holomorphic gauge. However, equation \eqref{Dterm} contains the adjoint $\Phi^\dagger$, which depends on the metric as
\be
\Phi^\dagger=H^{-1}\Phi^{+}H\,,
\ee
where the superscript $^+$ indicates complex conjugation and matrix transposition, and $H={\rm diag}(h,h^{-1})$. This brings a non-linearity in the partial differential equation \eqref{DiffD}, which can now be written as
\be\label{UnitaryDterm}
\Delta g = c + \frac{h_0^2\, |m|^2}{h^S}\, e^{2g}  \,,
\ee
where $h^S$, the determinant of the fixed hermitian metric on $S$, appears because of applying the Hodge-star operator on a four-form. This is a rather non-trivial equation that reduces to a Liouville-like equation when $m$ is constant and $h^S$ is the flat metric \cite{Cecotti:2010bp}. Nevertheless, there is a particularly nice setup in which  \eqref{UnitaryDterm} simplifies even further, as we discuss explicitly in the next subsection.

As a side remark, note that, for the split-type configurations \eqref{SplitV} we consider in this paper, the stability-based algebro-geometric criterion \cite{Simpson} for existence and unicity of solutions of the non-Abelian BPS equations \eqref{susy7} is trivially satisfied. For instance, it is immediate to see that  the only sub-bundle of $\mathbb{V}$ preserved by the Higgs field \eqref{NilpPhi} (i.e. $\L$) has negative $J$-slope, as enforced by the D-term equation \eqref{4dDterm}.

\subsection{The Hitchin Ansatz}

The most emblematic class of Higgs-bundle configurations is probably the one originally studied by Hitchin in the case of Riemann surfaces \cite{Hitchin:1986vp}. One can straightforwardly extend this Ansatz to the present context of complex surfaces, as first suggested in \cite{Vafa:1994tf}. This would correspond to taking the nilpotent Higgs field (\ref{NilpPhi}) such that the line bundle $\M$ is the trivial one, which amounts to demanding that\footnote{At weak coupling this is made compatible with cancellation of the Freed-Witten anomalies of the individual branes by considering a suitably-quantised primitive flux associated to the center-of-mass $U(1)$.}
\be\label{TopHitchin}
\mathcal{L}\simeq K_S^{-1/2}\,.
\ee
Since $S$ is compact, this choice implies that the quantity $m$ in \eqref{NilpPhi} can only be a constant. Notice also that equation \eqref{TopHitchin} only fixes the cohomology class of the gauge curvature in terms of that of $S$, but not its actual representative. Therefore, let us write the Ricci form of $S$ as
\be\label{Sigma}
\rho=\rho^{\h}-2i\del\delbar s({\bf x},{\bf \bar{x}})\,,
\ee
where $s$ is another globally well-defined smooth real function on $S$ such that $\int_S s\, \d {\rm vol}_S = 0$, and the factor of $2$ is for later convenience. Then, eq.\eqref{TopHitchin} states that $F^{\h}=\rho^{\h}/2$, or equivalently, using \eqref{Fsplit}, that\footnote{Recall that, in cohomology, $\tfrac{1}{2\pi}[\rho]=c_1(K_S^{-1})$.}
\be
F=\frac{\rho}{2}-i\del\delbar(g-s)\,.
\ee

Loosely speaking, $e^{g-s}$ is the conformal factor needed to rescale the hermitian metric on the surface $S$ to get the hermitian metric on the line bundle $\mathcal{L}$. More precisely we have
\be\label{ConfResc}
h_0=\sqrt{h^S}e^{-s}\,.
\ee
Using the above relation, our partial differential equation \eqref{UnitaryDterm} becomes
\be\label{HitchinVarphiU}
\Delta g = c +|m|^2e^{2(g-s)}\,,
\ee
where, as said, in this Hitchin set of solutions $m$ is a complex number. Let us now analyse two possible sub-cases of this setup.

\subsubsection*{K\"ahler-Einstein metric} 

The easiest possible situation is analogous to the one originally considered by Hitchin in the case of Riemann surfaces \cite{Hitchin:1986vp}. This arises when $g=s$. Taking into account the D-term condition \eqref{4dDterm}, which now simply says that $c=-|m|^2$, equation \eqref{HitchinVarphiU} reads
\be
\Delta g({\bf x},{\bf \bar{x}}) = 0\,,
\ee
whose unique solution on $S$ is $g({\bf x},{\bf \bar{x}})=0$. This, in turn, means that also $s=0$, and thus that both the gauge flux $F$ and the Ricci form $\rho$ are harmonic. If in particular $h^{1,1}(S) =1$, then $F_{\p}^{\h}=0$ in equation \eqref{SplitPrim} and therefore we have
\be
\rho=-\frac{|m|^2}{2}J\,.
\ee
Thus the metric on our surface $S$ is K\"ahler-Einstein with Einstein constant $-|m|^2/2$, that is it has constant negative Ricci curvature.

We can reverse the above argument and get a more useful statement. If we fix the metric on $S$ to be K\"ahler-Einstein, then $\rho=kJ$ with $k$ a real constant, which in particular means that $s=0$ in equation \eqref{Sigma}. Equation \eqref{UnitaryDterm} now reads
\be
\Delta g = |m|^2\left(e^{2g}-\frac{1}{{\rm Vol}(S)}\int_Se^{2g} \d{\rm vol}_S\right)\,,
\ee
where we substituted the value of $c$ fixed by the D-term \eqref{4dDterm}.
The above equation automatically implies that $g({\bf x},{\bf \bar{x}})=0$, because it admits a unique smooth solution. Therefore we conclude that, if we fix a (negatively curved) K\"ahler-Einstein metric on $S$, the vacuum solution for a constant nilpotent Higgs field involves a non-primitive, but still harmonic gauge flux.

\subsubsection*{Beyond K\"ahler-Einstein}

If instead we consider a non-K\"ahler-Einstein metric on $S$, the vacuum profile of the gauge flux will necessarily depart from the harmonic representative, and will be uniquely fixed by the equation
\be
\Delta g = |m|^2\left(e^{2g-2s}-\frac{1}{{\rm Vol}(S)}\int_Se^{2g-2s} \d{\rm vol}_S\right)\,.
\ee
As before, there will be a unique smooth solution for $g$. 
Note that this extension beyond K\"ahler-Einstein is also possible in the case of Riemann surfaces, thus directly generalising the type of solution discussed in \cite{Hitchin:1986vp}.

\subsection{Generalising the Ansatz}

There are a few ways of generalising the above simple set of solutions, namely by considering Higgs field profiles that are non-nilpotent and by considering line bundles $\CL$ that do not meet the topological condition (\ref{TopHitchin}). In the following we will consider and combine both generalisations, comparing the resulting equations for the function $g$ with the local T-brane solutions in the literature. 

\subsubsection*{Non-nilpotent Higgs field}

Let us first consider the case of four-cycles where the condition (\ref{TopHitchin}) is met, but now we have a non-nilpotent profile for the Higgs field. Namely we consider it to be of the form
\be\label{nNilpPhi}
\Phi=\left(\begin{array}{cc}
0&m\\ p &0
\end{array}\right)
\ee
where $p \in H^{2,0}(S,\CL^{-2})$, or equivalently a scalar holomorphic section of the line bundle $\P \equiv \mathcal{L}^{-2}\otimes K_S$. Notice that due to (\ref{TopHitchin}) we have that $\P \simeq K_S^2$. Such a bundle will have sections in many four-cycles of negative curvature, like for instance in those where $K_S$ also does. In this case eq. (\ref{UnitaryDterm}) generalises to
\be
\Delta g = c + h_S^{-1}  \left(|m|^2 h_0^2 e^{2g} - |p|^2 h_0^{-2} e^{-2g}\right)\,,
\ee
and so, using eq. (\ref{ConfResc}), we arrive to
\be\label{HitchinPainleve}
\Delta g = c + \left( |m|^2e^{2g} - h_0^{-4} |p|^2  e^{-2g} \right) e^{-2s}\, .
\ee
As before, $|m|^2$ is a constant, while $h_0^{-4} |p|^2$ is a globally well-defined smooth function on $S$. Finally, enforcing the 4d D-term condition implies that $c$ is given by
\be
c = - \frac{1}{{\rm Vol}(S)}\int_S \left(  |m|^2e^{2g} - h_0^{-4} |p|^2 e^{-2g} \right)e^{-2s}  \d{\rm vol}_S\, ,
\ee
so that eq. (\ref{HitchinPainleve}) has a (unique) solution. 

Notice that now $g$ will not vanish in the K\"ahler-Einstein case $s=0$. Instead, eq. (\ref{HitchinPainleve}) will become a complicated non-linear equation for $g$. Near the locus where $p=0$ we can Taylor expand the function $h_0^{-4} |p|^2$, and recover an equation very similar to that obtained in the local T-brane $\mathbb{Z}_2$ background of \cite{Cecotti:2010bp}. As pointed out in there, such an equation can be rewritten as a Painlev\'e III differential equation. Hence one would expect that, at least in a local patch near $p=0$, the profile for $g$ can be expressed in terms of solutions to that equation. Finally, one may depart from a K\"ahler-Einstein metric by considering $s \neq 0$. This will modify the (unique) solution for $g$, which will depend on the profiles of the functions $|m|e^{-s}$ and $h_0^{-2} |p|e^{-s}$. 

\subsubsection*{Non-trivial bundle $\CM$}

Let us now consider relaxing the topological condition \eqref{TopHitchin}, or in other words assume that $\CM \equiv \mathcal{L}^2\otimes K_S$ is a non-trivial bundle with sections. Given its definition, we can express the hermitian metric on $\CM$ as
\be
h_\CM = h_S^{-1} h_0^2\, e^{2g} = h_{\CM,0}\, e^{2(g-s)}\, ,
\ee
where $h_{\CM,0}$ corresponds to the metric with curvature $2F^{\h} - \rho^{\h}$ and $s$ is again defined by (\ref{Sigma}). We can then express (\ref{UnitaryDterm}) as 
\be
\label{NewUnitaryDterm}
\Delta g = c + \lVert m \rVert_\CM^2 \, e^{2(g-s)}\, ,  \qquad \qquad \lVert m \rVert_\CM^2 \equiv h_{\CM,0} |m|^2\,,
\ee
with $\lVert m \rVert_\CM$ a globally well-defined, smooth function on $S$ that vanishes over the same locus as $m$. This corresponds to an obvious generalisation of eq. (\ref{HitchinVarphiU}), where now the input function that determines $g$ is given by $e^{-s}\lVert m \rVert_\CM$. Since $\lVert m \rVert_\CM$ is non-constant, $g$ will be non-trivial even in the K\"ahler-Einstein case $s=0$, and so the gauge flux $F$ will depart from harmonicity.

Finally, one may combine a non-trivial bundle $\CM$ with a non-nilpotent Higgs field (\ref{nNilpPhi}), again assuming that $\P \equiv \mathcal{L}^{-2}\otimes K_S$ has sections. In that case, we may express the metric for this bundle as
\be
h_\P = h_S^{-1} h_0^{-2}\, e^{-2g} = h_{\P,0}\, e^{-2(g+s)}\, ,
\ee
with $h_{\P,0}$ the metric of curvature $-2F^{\h} - \rho^{\h}$. We then consider the globally well-defined, vanishing smooth function on $S$ given by $\lVert p \rVert_\P^2 \equiv h_{\P,0} |p|^2$. Together with the above definition for $\lVert m \rVert_\CM^2$, we obtain an equation for $g$ of the form
\be
\label{BHitchinPainleve}
\Delta g = c + \left( \lVert m \rVert_\CM^2 e^{2g} - \lVert p \rVert_\P^2  e^{-2g} \right) e^{-2s}\, .
\ee
While arising from a more general setup, this new differential equation is in fact very similar to (\ref{HitchinPainleve}), with the new functions that determine $g$ now given by $e^{-s} \lVert m \rVert_\CM$ and $e^{-s} \lVert p \rVert_\P$.

\section{A no-go theorem}
\label{sec:nogo}

The simple examples discussed in the previous section suggest that it is relatively easy to construct global T-brane configurations on four-manifolds with negative Ricci curvature. While it may seem that this preference comes from imposing the Hitchin Ansatz or generalisations thereof, there is in fact a deeper reason behind. Indeed, in the following we will see that compact T-brane configurations with Abelian gauge bundles cannot be implemented on four-manifolds of vanishing or positive Ricci curvature. We will first show this no-go result for the configuration with symmetry group $G = SU(2)$ and split gauge bundle of the type (\ref{SplitV}), and then generalise it to groups of higher rank. 

\subsubsection*{The case of SU(2)}

In order to investigate the possible obstructions to the construction of compact T-branes, let us first consider the stack of two D7-branes wrapping a simply-connected K\"ahler surface $S$, and with split gauge bundle $\mathbb{V}=\L\oplus\L^{-1}$. As before, we may start considering the T-brane background given by the nilpotent Higgs vev
\be\label{NilpotentPhiStack}
\Phi=\left(\begin{array}{cc}
0&m\\ 0 &0
\end{array}\right)\,,
\ee
where $m\in H^0(S,\M)$. Now, the very fact that an holomorphic section $m$ exists implies that the divisor associated to $\CM \equiv \mathcal{L}^2\otimes K_S$ is effective. That is, for $J$ in the K\"ahler cone we have
\be\label{ExistenceOfM}
\int_SJ\wedge c_1(\M)=\int_SJ\wedge(2c_1(\L)+c_1(K_S))\ge0
\ee
with the equality holding if and only if $\M$ is trivial.\footnote{We will always be at large volume, so in particular well away from boundaries of the K\"ahler cone.} Moreover, the 4d D-term condition (\ref{4dDterm}), or equivalently
\be\label{IntegratedDterm}
\int_S[\Phi,\Phi^\dagger]=-2\int_SJ\wedge F\,\cdot\sigma_3\,,
\ee
for a Higgs field of the form \eqref{NilpotentPhiStack} implies that
\be\label{NegFI}
2\int_SJ\wedge c_1(\L)<0\, ,
\ee
where we just used that $F/2\pi$ represents $c_1(\L)$ in cohomology. 
Subtracting the l.h.s. of \eqref{NegFI} to the middle expression in \eqref{ExistenceOfM}, we get the statement that we can construct such a T-brane in a region of K\"ahler moduli space where
\be\label{NonPosCurv}
\int_SJ\wedge c_1(K_S)>0\,.
\ee
This conditions forbids $S$ to be K3 or a manifold with positive-definite Ricci curvature. Indeed, if it were positive definite, the canonical class, which is represented by minus the Ricci form, would necessarily have a negative volume everywhere in K\"ahler moduli space. K\"ahler surfaces with negative-definite Ricci curvature certainly satisfy the necessary requirement \eqref{NonPosCurv}, but surfaces with indefinite curvature may also do so. The second inequality we get from \eqref{ExistenceOfM} and \eqref{NegFI} is
\be
\int_SJ\wedge c_1(\M)<\int_SJ\wedge c_1(K_S)\,,
\label{ineqnil}
\ee
which simply states that the volume of the holomorphic curve $\{m=0\}$ must be strictly smaller than the one of the self-intersection curve of $S$.\footnote{Note that such a curve needs not be holomorphic.} As a result, given a surface of non-positive curvature and a point in K\"ahler moduli space, \eqref{ineqnil} selects a subset of the lattice of bundles $[\CL]$ that one can use to build a T-brane background.

As an example, take the case where $S$ has only one K\"ahler modulus, i.e. $h^{1,1}(S)=1$. Together with the fact that $S$ is simply-connected, this implies that every gauge ``line bundle'' $\L$ on $S$ is of the form $\L\simeq K^{-n/2}$, for some non-zero integer $n$. Then, the two conditions \eqref{ExistenceOfM} and \eqref{NegFI} boil down to $n\leq1$ and $n>0$ respectively, which are both solved only by the choice $n=1$. This is nothing but the generalisation of Hitchin's class of solutions to a four-manifold, as already analysed in \cite{Vafa:1994tf} .

Let us now consider the most general Higgs vev compatible with a split rank-two gauge bundle, namely 
\be\label{NonNilpotentPhiStack}
\Phi=\left(\begin{array}{cc}
0&m\\ p &0
\end{array}\right)\,,
\ee
where now $m\in H^0(S,\M)$ and $p\in H^0(S,\P)$, with $\P \equiv \mathcal{L}^{-2}\otimes K_S$. Suppose now, without loss of generality, that the Fayet-Iliopoulos (FI) term in \eqref{IntegratedDterm} is positive, namely condition \eqref{NegFI} is satisfied. Then we obtain the following inequalities among the areas of the various curves involved
\be\label{OrderOfVolumes}
0\leq\int_SJ\wedge c_1(\M)<\int_SJ\wedge c_1(K_S)<\int_SJ\wedge c_1(\P)\,,
\ee
where again the first inequality (with equality if and only if $\M$ is trivial) comes from requiring that $\M$ admits at least one holomorphic section, as otherwise equation \eqref{IntegratedDterm} with positive FI term would be violated. Conversely, if the FI is negative, we get the same statement \eqref{OrderOfVolumes} with $\M$ and $\P$ swapped. In other words, the modes determining the sign of the D-term define the curve with the smallest volume. In any of these cases we have that (\ref{NonPosCurv}) must be satisfied, which again obstructs the construction of compact T-brane configurations on four-manifolds of vanishing or positive-definite Ricci curvature.

Incidentally, notice that the product $mp$ transforms as a section of $H^0(S,K_S^2)$, and it appears in the spectral equation for the Higgs field. Therefore for the background \eqref{NonNilpotentPhiStack} one could have guessed the obstruction to realise it on del Pezzo surfaces from a more standard, spectral-surface-based reasoning, see e.g. \cite{Hayashi:2010zp}. Nevertheless, our analysis provides more detailed information about the obstruction, like for instance the inequalities \eqref{OrderOfVolumes} that select a subset of possible line bundles  $[\CL]$.

\subsubsection*{Higher rank groups}

Let us now consider a general simple Lie group $G$, of Lie algebra $\mathcal{G}$ specified by a Cartan subalgebra $H_i$ and the set of roots $E_{\rho}$. In the canonical basis, they satisfy the following set of relations
\be
\begin{array}{ccc}
\ [H_i, E_\rho] & = & \rho^i E_\rho\\ 
\ [E_\rho ,E^\dagger_\rho] &=& \sum_i \rho^{i} H_i
\end{array}
\qquad\qquad i =1,\ldots, {\rm rank}(\mathcal{G})\, .
\label{canonical}
\ee 
For our purposes it is more convenient to instead consider the algebra in the so-called Chevalley basis. The latter is specified with respect to a chosen set of simple roots:
\be\label{Chevalley}
\begin{array}{ccc}
\ [h_i,e_j]&=&C_{ji}e_j \\ 
\ [e_i,e^\dagger_j]&=&\delta_{ij} h_j
\end{array}
\qquad\qquad i,j=1,\ldots, {\rm rank}(\mathcal{G})\,,
\ee
where $h_i$ are the Cartan generators and $e_i$ the generators associated to the simple roots in this basis. Finally, $C_{ij}$ the Cartan matrix, that can always be decomposed as
\be
C \, =\, D \, S\, , \qquad \qquad D_{ij} = \frac{2\delta_{ij}}{\a_j \cdot \a_j}\, , \qquad \qquad  S_{ij} = \a_i \cdot \a_j \, ,
\ee
where $\a_i$ stand for the simple-root vectors in the canonical basis \eqref{canonical}. There, a general root vector can be decomposed as
\be
\rho \, =\, \sum_i v_{\rho}^i \a_i \qquad \qquad v_{\rho}^i \in \IZ \, ,
\ee
and then for its corresponding generator in the Chevalley basis we have that
\be
[h_i, e_{\rho}] =  q_{\rho}^i  e_{\rho}\, ,  \qquad \qquad q_{\rho}^i = \sum_j v_{\rho}^j C_{ji}\, .
\ee

In this setup, let us take the following Ansatz for our T-brane background
\be
\frac{F}{2\pi} = \sum_i  \omega_i h_i = \sum_i c_1(\CL_i)h_i
\ee
and
\be
\Phi = \sum_{\g \in R'} m^\g e_\g\, ,
\ee
where $m^\g \in H^{2,0}(\otimes_i (\CL_i)^{q_{\g}^i})$ and $\g$ runs over a root subset $R'$ such that 
\be
[e_\g, e_{\b}^\dag] =  \delta_{\g\b} \sum_i \g^i h_i\, , \qquad \forall \g, \b \in R'\, .
\ee
As a result we have
\be
[\Phi, \Phi^\dag] = \sum_{\g, i}  m^\g \wedge \bar{m}^\g\, \sig_{\g}^i \,  h_i \, ,
\ee
with
\be
\sig_{\g}^i \, =\, \sum_j D_{ij} v_{\g}^j \, .
\ee

Given this background, the fact that $m^\g$ are holomorphic sections implies
\be
\int_S \left( \sum_i q_{\g}^i \, c_1(\CL_i) + c_1(K_S) \right) \wedge J \ge 0 \quad \quad \forall \g \in R'\, .
\label{effective}
\ee
In addition, the D-term condition implies that
\be
\int_S c_1(\CL_i) \wedge J = - \sum_\g \sig_{\g}^i \, \| m^{\g}\|^2
\ee
where we have defined
\be
\| m^{\g}\|^2 \equiv \oh  \int_S m^{\g} \wedge \bar{m}^{\g}\, .
\ee

Therefore
\be
\sum_i q_{\g}^i \int_S c_1(\CL_i) \wedge J = - \sum_{i, \b} q_{\g}^i \,  \sig_{\b}^i \,  \| m^{\b}\|^2 =  - \sum_{\b \in R'} v_\g^t\, D S D\, v_\b \,  \| m^{\b}\|^2
\qquad \forall \g  \in \, R'.
\ee
Now, notice that the matrix 
\be
A_{\g\b} \, =\,  v_\g^t \, D S D \, v_\b \, =\, \sig_\g^t \, S \, \sig_\b
\ee
is semi-definite positive, and definite positive when the set of vectors $\{v_\g\}$, $\{\sig_\g\}$ or $\{q_\g\}$, $\g \in R'$ are linearly independent. Therefore, when $\{v_\g\}$ are not linearly independent there are zero modes of $A_{\a\b}$ that correspond to D-flat directions.\footnote{Moreover, in this case one is able to form a product of sections of the form
\be
m^{\g_1}m^{\g_2}\dots m^{\g_n} \in H^0(K_S^n)
\ee
which cannot exists in a positive curvature four-cycle. Therefore, in positive curvature four-cycles one can consider the $\{\rho_\a\}$ to be linearly independent} Going along them one can switch off the necessary number of vevs in the subset of roots $R'$ such that it gets reduced to $R''$, that corresponds to a set of linearly independent vectors. For this new subset $R''$ we have that $A_{\g\b}$ is positive definite, and then we have that
\be
\sum_{\g, i}  \| m^{\g}\|^2 \int_S q_{\g}^i \, c_1(\CL_i) \wedge J = - \sum_{\g, \b} A_{\g\b} \| m^{\g}\|^2 \| m^{\b}\|^2\, < \, 0
\ee
where now $\g, \b \in R''$. As a result
\be
\int_S c_1(K_S) \wedge J  > \int_S \left( \frac{\sum_{\g, i}  \| m^{\g}\|^2 q_{\g}^i c_1(\CL_i)}{\sum_\g \| m^{\g}\|^2}  + c_1(K_S) \right) \wedge J \geq 0 \, .
\ee
where in the second inequality we have made use of \eqref{effective}. Notice that when we have only one $\g$ this equation reduces to 
\be
\int_S c_1(K_S) \wedge J  > \int_S \left( \sum_i q_{\g}^{i} c_1(\CL_i)  + c_1(K_S) \right) \wedge J > 0
\ee
familiar from the $SU(2)$ case.

\section{T-branes and stability  walls}\label{sec:stability}

Starting from a T-brane configuration, we now want to study its stability when we move in the moduli space of K\"ahler structures. Changes are expected to arise simply because the r.h.s. of the D-term equation \eqref{IntegratedDterm} depends on the K\"ahler form. In particular, if $S$ has more than one K\"ahler modulus, there will generically be real codimension-one loci in the K\"ahler moduli space where the r.h.s vanishes, possibly resulting in a decay of the T-brane, or in its transmutation into a different type of supersymmetric vacuum. In this section, we would like to make a systematic study of what may happen to the T-brane background as we cross such stability walls. We will first consider the sort of T-brane configurations considered in section \ref{sec:compact}, and then extend our analysis to a system of two D7-branes intersecting at a curve. 

\subsection{Coincident branes}\label{coincbranes}

Let us consider two D7-branes wrapping a simply-connected K\"ahler surface $S$, holomorphically embedded in a Calabi-Yau threefold. As in section \ref{sec:compact} we consider a split rank-two gauge bundle of the form (\ref{SplitV}), specified by a line bundle $\mathcal{L}$ of curvature $F$. We moreover consider a K\"ahler structure compatible with a T-brane of the nilpotent type \eqref{NilpotentPhiStack}. Because of the D-term \eqref{IntegratedDterm}, the size of the vev $\langle m\rangle$ is controlled by the FI term $\int F\wedge J$, and thus it is proportional to the distance from the wall, which is defined by the condition $\int F\wedge J =0$. There we get a vanishing vacuum expectation value for $\Phi$ and therefore a standard system of two coincident D7-branes with a worldvolume flux along the Cartan. We are now interested in studying the open-string moduli space in a region around the origin 
\be
\Phi =0\,,
\ee
and to see how the D7-brane system evolves when the FI term is switched back on, at the other side of the wall.

To carry such an analysis one may first consider the spectrum of light open-string modes at the wall, where the effective theory has a $U(1)\times U(1)$ gauge group and a set of bifundamental chiral fields charged under the relative $U(1)$, associated to the Cartan.
By standard results \cite{Katz:2002gh} (see also \cite{Blumenhagen:2008zz}), the full spectrum of charged massless fields is provided by the appropriate sheaf extension groups. More precisely, as in section \ref{sec:compact}, let us define the two line bundles $\M\equiv\L^2\otimes K_S$ and $\P\equiv\L^{-2}\otimes K_S$, with $K_S$ the canonical bundle of $S$. Then one has
\bea\label{SpecCoinc}
(+)&\in&{\rm Ext}^1(i_*\L^{-1},i_*\L)\simeq \underbrace{H^0(S,\M)}_{m}\;\oplus\;\underbrace{H^1(S,\P)}_{a_+}\,,\nonumber \\ \\
(-)&\in&{\rm Ext}^1(i_*\L,i_*\L^{-1})\simeq \underbrace{H^0(S,\P)}_{p}\;\oplus\;\underbrace{H^1(S,\M)}_{a_-}\,,\nonumber
\eea
where the signs on the left indicate the relative-$U(1)$ charge and $i$ is the embedding map of $S$ in the Calabi-Yau threefold. Here the $H^0$ parts correspond to massless off-diagonal fluctuations of the Higgs field, whereas the $H^1$ parts correspond to off-diagonal components of the non-Abelian gauge field living on $S$. Notice that a non-vanishing vacuum expectation value for the latter would correspond to a non-Abelian gauge bundle, and so the vevs for such fields $a_\pm$ were assumed to vanish in the T-brane configurations of section \ref{sec:compact}. We must however take them into account in the following, to study how the D-brane configuration may react as we cross a stability wall.

On top of the charged modes there are also uncharged zero modes, which however only appear as fluctuations of $\Phi$ and not of the gauge field, because we are taking $S$ to be simply-connected. Such fields originate from open strings with endpoints on the same D7-brane and thus corresponding to its normal deformations inside the ambient Calabi-Yau manifold. Here we only focus on relative deformations of the two branes wrapping $S$, and ignore the movements of their center of mass. Therefore, these deformations appear in the Higgs-field fluctuation as
\be
\label{neutral}
\delta\Phi|_{\rm neutral}=\left(\begin{array}{cc}
v&0\\ 0 &-v
\end{array}\right)\,,\qquad\quad v\in H^0(S,K_S)\,.
\ee
Note that these vevs were also set to vanish in the T-brane configurations of section \ref{sec:compact}.

Finally, the absence of modes with negative norm (ghosts) for the strings connecting the two branes \cite{Donagi:2008ca} leads to the following important requirements
\be\label{NoGhosts}
H^0(S,\L^2)=H^0(S,\L^{-2})=0\,.
\ee
These conditions are automatically satisfied if the FI term vanishes and we are inside the K\"ahler cone. 

Given the above spectrum one may analyse how the system behaves at both sides of the wall. For simplicity, we will first consider the case where the modes (\ref{neutral}) are absent. Then, in a sufficiently small region in K\"ahler moduli space around the wall, and upon dimensional reduction to 4d, the D-term condition (\ref{Dterm}) becomes\footnote{We use the same symbol for the eight-dimensional fields and the corresponding four-dimensional zero modes, and suppress the symbol $\langle\cdot\rangle$ to indicate the vev. We moreover work in units of $\alpha'$.}
\be\label{OtherSide}
\sum_m|m|^2+\sum_{a_+}|a_+|^2-\sum_p|p|^2-\sum_{a_-}|a_-|^2=\xi\,,
\ee
which is nothing but the vanishing of the 4d D-term scalar potential. By assumption, on one side of the wall we have a supersymmetric configuration where only $m$-type zero modes have a non-vanishing vev, and so there $\xi >0$. Then we reach the wall by moving in the K\"ahler-structure moduli space. After crossing the wall the FI term flips sign, so
\be
\xi\equiv-2\int_S J\wedge c_1(\L)<0\,.
\ee
Therefore from equation \eqref{OtherSide} it is manifest that if $H^0(S,\P)=H^1(S,\M)=0$, there is no solution for the D-term equation as we cross the wall. Microscopically, this means that the T-brane we started with disappears as we cross the wall, by decaying into its D7-brane constituents, which are not mutually supersymmetric.\footnote{Note that we are considering the D7-brane stack in isolation, neglecting other D-branes that may yield further chiral zero modes charged under the Cartan $U(1)$. One clearly needs to take into account the full brane content of the compactification to see if crossing the wall really breaks supersymmetry.}

Interestingly, by using the index theorem we are able to formulate a practical \emph{necessary} criterion for such a decay to occur.
In particular, applying the index theorem to the line bundle $\P$, we get
\be\label{IndexP}
h^0(S,\P)-h^1(S,\P)=\int_S {\rm ch}(\P)\wedge {\rm Td}(S)\,,
\ee
where the symbol $h^i$ indicates the dimension of the corresponding group $H^i$, ``${\rm ch}$'' is the total Chern character and ``${\rm Td}$'' is the Todd class.\footnote{For a line bundle $\mathcal{F}$, ${\rm ch}(\mathcal{F})=1+c_1(\mathcal{F})+c_1^2(\mathcal{F})/2$, and for a surface $S$ one has ${\rm Td}(S)=1-c_1(K_S)/2+(c_1(K_S)^2+c_2(S))/12$.} In \eqref{IndexP} we have used that $h^2(S,\P)=h^0(S,\L^2)=0$, where the first equality comes from Serre duality, and the second from equation \eqref{NoGhosts}. Likewise, the index theorem for the line bundle $\M$ means that
\be\label{IndexM}
h^0(S,\M)-h^1(S,\M)=\int_S {\rm ch}(\M)\wedge {\rm Td}(S)\,,
\ee
where again we used that $h^2(S,\M)=h^0(S,\L^{-2})=0$, because of Serre duality and equation \eqref{NoGhosts} respectively. By subtracting equation \eqref{IndexM} to equation \eqref{IndexP}, with some trivial algebra we get to the chiral index of the theory:
\be\label{Index}
I = \#(+)-\#(-)=2\int_S c_1(\L)\wedge c_1(K_S)\,,
\ee
where the symbol $\#(\pm)$ denotes the number of zero modes with $U(1)$-charge $\pm$. Finally, from equation \eqref{Index} we obtain the following implication
\be
I = 2\int_S c_1(\L)\wedge c_1(K_S) \; \leq \; 0\qquad \Longrightarrow\qquad {\rm No\;\; T{\rm-}brane\;\; decay}\,,\label{TbraneNecCond}
\ee
because if there were no negatively-charged modes available to turn the T-brane into another supersymmetric system, the integral on the l.h.s. would necessarily be positive.

On the contrary, if conditions are met for some negatively-charged modes to exist, the T-brane simply turns into a different supersymmetric state on the other side of the wall.\footnote{One particular case is when $I=0$, which in the literature corresponds to a wall of threshold stability. Indeed, by looking at the definition (\ref{Index}) one realises that $-I$ corresponds to the intersection product used in \cite{deBoer:2008fk} to classify stability walls.}  The latter could be another T-brane, if just the $p$-type modes get a vev, a non-Abelian bundle configuration (T-bundle) if just the $a_-$-type modes get a vev, or a more complicated mixed object. The indices of the individual bundles, quoted in equations \eqref{IndexP} and \eqref{IndexM}, can turn useful to guess what type of object the T-brane may turn into, although most of the times they cannot give definite answers. In practice, one may compute the cohomology groups in \eqref{SpecCoinc} case by case, as illustrated in appendix \ref{ap:examples}, to find out the fate of the T-brane at the other side of the wall. There are however a few classes of constructions where a more general statement can be made, as we discuss in the following.

\subsubsection*{The Hitchin Ansatz}

An interesting case of T-branes is the one constructed using what we have dubbed the Hitchin Ansatz, namely when $\CM$ is trivial, or equivalently $\L\simeq K_S^{-1/2}$. One important remark regarding this case is that, if the Ricci curvature of $S$ is negative definite, then there will be no stability walls. Indeed, for $\L\simeq K_S^{-1/2}$ we have that the FI term becomes
\be
\xi=\int_S J\wedge c_1(K_S)\,,
\ee
which for negative curvature cannot be taken to zero while moving inside the K\"ahler cone.

Let us then consider the case where the Ricci curvature of $S$ is indefinite. This in particular implies absence of holomorphic sections for the canonical bundle (thus $S$ is rigid) and for any power thereof (positive and negative). Therefore no $p$-type modes are available and, since by assumption $S$ is simply-connected, no $a_-$-type modes are available either. Hence, in this class of configurations, our T-brane is forced to decay into a non-supersymmetric vacuum when the wall is crossed.

A simple instance of a K\"ahler surface with the above properties can be obtained as follows. Consider $\mathbb{P}^4$ with homogeneous coordinates $x_1,\ldots,x_5$, blown up along a four-cycle, e.g. $\{x_1=x_2=0\}$. The toric weights of this manifold are
\be
\begin{array}{cccccc}
x_1 & x_2 & x_3 & x_4 & x_5 & w  \\ \hline
1 & 1 & 0 &0 &0 & -1\\
0 & 0 & 1 &1 &1 & 1
\end{array}
\ee
where $E:\{w=0\}$ corresponds to the exceptional divisor, homeomorphic to $\mathbb{P}^2\times\mathbb{P}^1$. In this ambient manifold, we consider the Calabi-Yau threefold CY$_3$ given by the zero-locus of a smooth polynomial of bi-degree $(1,4)$, and the D7-brane stack wrapped on $S:E\cap{\rm CY}_3$. It is easy to show that this surface is rigid (as a consequence of the rigidity of the exceptional divisor), and moreover has indefinite Ricci curvature, because e.g.
\be
\int_{S\cap\{x_1=0\}} c_1(K_S)=4\,,\qquad\;\int_{S\cap\{x_3=0\}} c_1(K_S)=-3\,.
\ee
By using the Hirzebruch-Riemann-Roch theorem, we can also easily show that this surface has no cohomologically non-trivial one-forms
\be
h^{0,1}(S)=1-\frac{1}{12}\int_Sc_1^2(K_S)+c_2(S)=0\,,
\ee
where we used that $h^{0,2}(S)=0$. If we label $H:\{ x_1 = 0 \}$ and expand the K\"ahler form in this basis, $J \equiv v_H \, H + v_E \, E$, we may compute the Fayet-Iliopoulos term as $\xi = 5 ( 4 v_H - 7 v_E )$, which can indeed acquire both positive and negative values within the K\"ahler cone.

\subsubsection*{Negative curvature}

Let us now consider the case where the Ricci curvature of the surface $S$ is negative definite. Note that this does not necessarily imply that $S$ can be holomorphically deformed, a subcase to be considered momentarily. By the observation made above, in the negative curvature case we must consider a T-brane whose $m$-type mode transforms under a non-trivial bundle $\CM$. The fact that $\CM$  is effective and non-trivial, together with the ampleness of $K_S$ due to the negative curvature, implies that
\be
\label{yesmmode}
I > -\int_S c_1^2(K_S)\,,
\ee
where the r.h.s. is a negative integer number.  Applying the same reasoning to the bundle $\P$, we have that the existence of $p$-type modes implies that $I \leq \int_S c_1^2(K_S)$, and so whenever
\be
\label{nopmode}
I > \int_S c_1^2(K_S) > 0
\ee
there will be no such $p$-modes. Notice that  imposing (\ref{nopmode}) implies (\ref{yesmmode}). Therefore, if we consider a case where (\ref{nopmode}) is satisfied and $h^1(S,\M)=0$ (see appendix \ref{ap:examples} for an example), then there will be a T-brane decay. Alternatively, if $h^1(S,\M)>0$ then the T-brane will turn into a supersymmetric non-Abelian bundle configuration on the other side of the wall.

One particular case of a negative curvature four-cycle is when $S$ can be holomorphically deformed, namely when the modes (\ref{neutral}) exist. Then there is a self-intersection curve defined by $\C \equiv \{v = 0\}$ and with a genus $g$ such that
\be
\int_S c_1^2(K_S) = g - 1\, .
\ee
Note that by the adjunction formula one finds that $g=1+[S]^3$, where $[S]$ stands for the divisor class of $S$ in the Calabi-Yau. Since $\int_S c_1^2(K_S)>0$, we have that $[S]^3$ is a positive number and so $g \geq 2$.

In this particular case there is the open-string field $v$ defined in (\ref{neutral}), which is a modulus along the wall. One may then wonder what happens when the wall is crossed with a non-vanishing Higgs-field vev, namely at 
\be
\label{vevPhi}
\Phi =\left(\begin{array}{cc}
v&0\\ 0 &-v
\end{array}\right)\,.
\ee
In this case, by dimensionally reducing the D7-brane superpotential
\be
W\, =\, \int_S \tr \left(\mathbb{F} \wedge \Phi\right)\,,
\ee
one obtains Yukawa couplings of the form 
\be
W \supset d_{ijk}\, v^i a_-^j a_+^k\,,
\ee
which generically give an F-term mass to the negative-chirality modes $a_-$. Now, if we impose (\ref{nopmode}) and cross the wall at (\ref{vevPhi}), for $h^1(S,\M)>0$ there will be an F-term potential that will make (\ref{vevPhi}) vanish and take the system to the supersymmetric configuration of coincident D7-branes with a non-Abelian bundle created by the vev of $a_-$.

Notice that at (\ref{vevPhi}) we have a system of two homotopic D7-branes intersecting at a curve $\C$, with opposite worldvolume fluxes. This is nothing but a particular case of a more general configuration, made of two intersecting D7-branes with arbitrary worldvolume fluxes. As we will now see, one can formulate the T-brane wall-crossing conditions for this more interesting case as well. 

\subsection{Intersecting branes}

Let us consider two D7-branes wrapping different simply-connected K\"ahler surfaces $S_1,S_2$, holomorphically embedded in a Calabi-Yau threefold. Let $\L_1,\L_2$ be the holomorphic gauge line bundles on each of the two branes, with fluxes $F_1,F_2$ respectively. As in the coincident case, the four-dimensional effective theory has a $U(1)\times U(1)$ gauge group and bifundamental chiral fields charged under the relative combination. The 4d D-term condition that controls the vacuum expectation values of their scalar components is now given by
\bea\label{GeneralD}
\sum_{m\in(+,-)}|m|^2-\sum_{p\in(-,+)}|p|^2 &=& \int_{S_2}J\wedge F_2-\int_{S_1}J\wedge F_1 \, =\, \xi\, ,
\eea
where the two sums extend over zero modes with $U(1)\times U(1)$-charges $(+,-)$ and $(-,+)$ respectively. They correspond to open strings stretching from brane $2$ to brane $1$ and to strings going the opposite way respectively. Assuming that the intersection curve $\C\equiv S_1\cap S_2$ is connected, such zero modes are counted by the following sheaf extension groups \cite{Katz:2002gh} (see also \cite{Blumenhagen:2008zz}):
\bea\label{SpecInters}
(+,-)&\in&{\rm Ext}^1(i_{2*}\L_2,i_{1*}\L_1)\simeq H^0(\C,\L_2^{-1}|_{\C}\otimes\L_1|_{\C}\otimes K_{\C}^{1/2})\,,\nonumber \\ \\
(-,+)&\in&{\rm Ext}^1(i_{1*}\L_1,i_{2*}\L_2)\simeq H^0(\C,\L_2|_{\C}\otimes\L_1^{-1}|_{\C}\otimes K_{\C}^{1/2})\,,\nonumber
\eea
with $K_{\C}$ its canonical bundle, and $i_1,i_2$ the embedding maps of branes $1,2$ respectively.

In this case the wall is defined by the K\"ahler structure slice where $\int F_1\wedge J=\int F_2\wedge J$. There we have a system of two intersecting D7-branes, and thus the spectrum of massless fluctuations is given by equation \eqref{SpecInters}. Notice that, unlike in the coincident case, now the spectrum of zero modes is only counted by modes of the Higgs field. We now assume that there is at least one of these two types of modes, say a $m$-type mode with charge $(+,-)$, so that, at one side of the wall ($\xi > 0$), there is a supersymmetric bound state with a T-brane profile localised at $\C$.  As we cross the wall to the other side, either this T-brane turns into a different kind of T-brane or, if no $p$-type mode is available, the T-brane decays into the two mutually non-supersymmetric constituents.\footnote{This decay process has been discussed in \cite{Collinucci:2014qfa}.}

\begin{figure}[htb]
\begin{center}
\includegraphics[width=100mm]{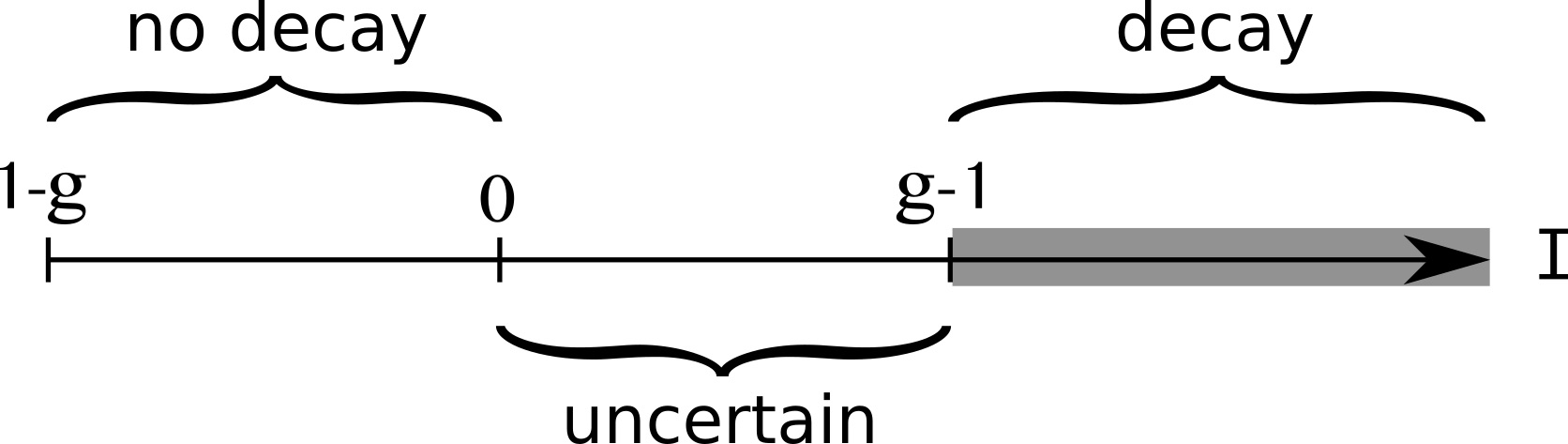}
\caption{Different possibilities of decay into non-BPS constituents as a T-brane constructed from two intersecting D7-branes crosses a stability wall.}\label{decay}
\end{center}
\end{figure}

Since in this case the spectrum of charged zero modes is simpler, we are able to formulate a \emph{sufficient} criterion for our T-brane to decay across the wall. First, notice that the chiral index of the theory is given by 
\be
I \equiv {\rm deg}\,\L_1|_{\C}-{\rm deg}\,\L_2|_{\C}\; = \frac{1}{2\pi} \int_{\C} F_1 - F_2 \,.
\ee
Let us for now assume that the surfaces $S_1$, $S_2$ do not have holomorphic deformations or, if they do, that none of them will split the intersection curve into multiple connected components. Then, calling $g$ the genus of $\C$ and using the Riemann-Roch theorem, the existence of the $m$-type mode we began with implies that
\be\label{ExistenceOfmInters}
I \;\geq\; 1 - g\,,
\ee
with the equality holding if and only if $m$ is constant, which is the analogue of the Hitchin Ansatz for a system of intersecting D7-branes. This relation comes from the fact that the degree of a line bundle on a curve coincides with the number of zeros minus the number of poles of any of its rational sections. Moreover, we have the analogue of \eqref{TbraneNecCond}, with the index theorem adapted to this case
\be
I  \leq \;0\qquad \Longrightarrow\qquad {\rm No\;\; T{\rm-}brane\;\; decay}\,.\label{TbraneNecCondInt}
\ee
Finally, by the same reasoning, if the condition
\be\label{Kodaira-Intersecting}
I \;>\;g - 1
\ee
is satisfied, there are no $p$-type modes to form a T-brane on the side of the wall where the FI term is negative. Therefore, we readily see that, if the two D7-branes intersect on a sphere, the fate of our T-brane is to decay when we cross the wall. The same statement holds true when $\C$ is a two-torus and $\int_\C F_1 \neq \int_\C F_2$. We therefore obtain a simple picture for the decay possibilities of intersecting D7-branes, summarised in figure \ref{decay}.

%
%
If on the other hand the surfaces $S_1$, $S_2$ contain holomorphic deformations such that $\C$ splits into multiple components, the wall-crossing picture just described may change. Indeed, when the matter curve $\C= \cup_a \C_a$ is disconnected, one needs to apply \eqref{SpecInters} separately to each individual component $\C_a$ to obtain the massless spectrum. While then the relations \eqref{ExistenceOfmInters} and \eqref{TbraneNecCondInt} continue to hold,\footnote{More precisely, \eqref{ExistenceOfmInters} should be written in terms of topological invariants as $I\geq h^{0,0}(\C)-h^{0,1}(\C)$.} the sufficient condition for decay \eqref{Kodaira-Intersecting} gets replaced by a significantly weaker one. This is because it is enough to find at least a $p$-mode localised on any of the connected components of $\C$, in order for the two branes to bind back again into a supersymmetric system across the wall. In other words, decay will only occur when all the available holomorphic deformations of $S_1$ and $S_2$ split $\C$ in such a way that on every component $\C_a$ one has $I_a>g_a-1$.

\section{Conclusions}
\label{sec:conclu}

In this paper we have analysed global aspects of T-branes in type IIB/F-theory compactifications. In this context T-branes were first presented as interesting configurations that allow for hierarchical Yukawas in F-theory GUTs. Since the computation of Yukawas can be essentially done within a local patch of the four-cycle $S_{\rm GUT}$, only a local description of the T-brane background is needed to realise this property. Nevertheless, this local picture inevitably misses some crucial features of T-branes, including possible obstructions to their existence, that can only be revealed by a global analysis.

In this spirit we have given a global description of such T-brane configurations from the viewpoint of the K\"ahler four-cycle $S$ where they are defined. We have focused on T-branes with a pole-free holomorphic Higgs field $\Phi$, and an Abelian gauge flux $F$, which we have dubbed compact T-branes. We have observed several general features that mainly depend on the topology of $S$ and the pull-back of the threefold K\"ahler form $J$. Namely we have found that:

\begin{itemize}

\item[-] In general, the worldvolume flux $F$ lies in a non-harmonic representative of its cohomology class. The departure from harmonicity is codified in a globally well-defined function $g$ on $S$ satisfying certain non-linear PDEs. In local patches, such equations reproduce the ones already found in the T-brane literature.

\item[-] There is an obstruction to building these T-brane backgrounds on surfaces where the Ricci curvature class vanishes or is positive definite. In the remaining surfaces the existence of T-branes depends on the classes $[\rho], [F] \in H^2(S)$ of the Ricci form and the worldvolume flux, respectively, as well as on the point in K\"ahler moduli space. For instance, in the simplest case, the following condition needs to be satisfied:
\be\label{ConcuVol}
0\leq\int_SJ\wedge (2F-\rho) < - \int_SJ\wedge \rho \, .
\ee
Hence, given a four-cycle $S$ and a point in K\"ahler moduli space, only the subset of quantised fluxes $F$ satisfying \eqref{ConcuVol} will be suitable to construct a compact T-brane. Notice that whenever the Ricci form has a negative sign when projected into the K\"ahler form, one may choose $[F] = [\rho]/2$ (i.e. the Hitchin Ansatz) to satisfy \eqref{ConcuVol}. 

\item[-] In those regions of K\"ahler moduli space where $0 < \xi \alpha' = -\tfrac{1}{\pi\alpha'} \int_S F\wedge J \ll 1$, we may interpret our T-brane background as a 7-brane bound state obtained after switching on a Fayet-Ilioupoulos term $\xi$, and see the slice $\xi=0$ as a T-brane stability wall. The fate of the system as the wall is crossed to the region $\xi <0$ again depends on the T-brane topological data, and in particular on the two classes $[\rho]$ and $[F]$. A similar statement holds for a T-brane built at the intersection of two 7-branes. 

\end{itemize}

These general results already suggest many avenues for further investigation. The two most pressing questions are perhaps {\it i)} how everything generalises when we allow for T-brane systems with poles, and {\it ii)} what are the implications of our findings for concrete F-theory GUT models. We may for instance consider a model where $S_{\rm GUT}$ hosts an exceptional symmetry group like $G=E_{6,7,8}$ and a T-brane sector within a subalgebra of $G$, as it is the case for local models of Yukawas \cite{Cecotti:2010bp,Chiou:2011js,Font:2013ida,Marchesano:2015dfa,Carta:2015eoh}. Then our no-go result implies that either {\it a)} $S_{\rm GUT}$ cannot be del Pezzo or {\it b)} the T-brane sector contains some poles. In the latter case, one might interpret such poles as being sourced by further 7-branes intersecting $S_{\rm GUT}$ on matter curves, and it would be interesting to engineer compactifications that reproduce such a setup.  

An additional generalisation would be to look at T-brane backgrounds where the gauge bundle is not of the split form \eqref{SplitV}. One simple way of obtaining non-split bundles is by switching on any of the bundle moduli $a_+, a_-$ in \eqref{SpecCoinc} on top of a T-brane background near the stability wall. Obviously, the no-go result of section \ref{sec:nogo} still holds for these more complicated configurations. In general, for any non-split bundle that can be taken to the split form by moving in open-string moduli space the no-go result will apply, and equation \eqref{NonPosCurv} should be satisfied. It would be therefore very interesting to analyse the structure of the open-string moduli space around general T-brane backgrounds. 

Another direction would be to examine how $\alpha'$ corrections modify the T-brane constructions considered in this paper. At moderate volumes of the compactification one may in principle apply the same strategy as in \cite{Marchesano:2016cqg} to see how such corrections affect the differential equations of section \ref{sec:compact}, that govern the 7-brane background. However, as these corrections do not affect the holomorphic T-brane data and are sufficiently mild not to flip the FI-term sign, the no-go theorem of section \ref{sec:nogo} should still hold. 

Finally, as the necessary conditions for the existence of compact T-branes depend on the point in the K\"ahler moduli space of the compactification, it would be interesting to see if our results could have any implications for K\"ahler moduli stabilisation.

In summary, as argued in the introduction, our findings can be seen as one further step in the classification of the full set of BPS branes in type IIB/F-theory compactifications. As such, they should have direct consequences  for the model-building applications that triggered the recent study of T-branes in this context, and it would be interesting to fully explore such implications. In any event, we expect that having a good understanding of global T-brane configurations will give rise to new insights in the comprehension of string theory vacua.

\bigskip

\centerline{\bf \large Acknowledgments}

\bigskip

We would like to thank Luis \'Alvarez-C\'onsul, Mario Garc\'ia-Fern\'andez, Cumrun Vafa, Taizan Watari and Timo Weigand for useful discussions. R.S. wants to thank Andr\'es Collinucci for collaboration on related topic and many useful conversations.
This work is supported by the Spanish Research Agency (Agencia Estatal de Investigaci\'on) through the grant IFT Centro de Excelencia Severo Ochoa SEV-2016-0597, by the grant FPA2015-65480-P from MINECO/FEDER EU, and the ERC Advanced Grant SPLE under contract ERC-2012-ADG-20120216-320421. S.S. is supported by the FPI grant SVP-2014-068525.


\appendix

\section{4d interpretation of flux non-harmonicity}\label{ap:alpha}

In section \ref{ss:global} we defined $\d\a = - i\partial\bar{\partial}g$ to be the exact part of the worldvolume flux that typically appears in T-brane solutions. For intersecting branes, a non-harmonic exact flux profile would break supersymmetry, and it would be seen as turning a non-vanishing vev for a Kaluza-Klein mode for the gauge vector field. If we consider a T-brane in the vicinity of a stability wall of the sort analysed in section \ref{coincbranes}, this correspondence between non-harmonic fluxes and Kaluza-Klein modes remains to a good extent accurate. Therefore, it is natural to interpret $\a$ as a set of KK modes that got a vacuum expectation value when the 4d Fayet-Iliopoulos term was switched on and the system evolved to a T-brane background. In the following we would like to give a more precise description of this intuition, in terms of the 4d effective gauge theory. 

Let us begin with the D-term part of the 8d action, which is given by \cite{Beasley:2008dc}
\begin{align} 
S &\supset \int_{\mathbb{R}^{1,3} \times S} \, \rm{Tr} \left( \mathcal{D} \wedge * \mathcal{D} \right) \label{Daction}\\
\mathcal{D} &= - * \left( J \wedge F + \frac{1}{2} [ \Phi, \Phi^\dagger ] \right) \\
&= * \left( - \frac{c}{4} J \wedge J - J \wedge \rm{d} \alpha - \frac{1}{2} * \varphi \right) \, \sigma_3 \nonumber,
\end{align}
where we have applied the general Ansatz of section \ref{coincbranes} and in particular made use of eqs. \eqref{varphi} and \eqref{SplitPrim}. To convert this to a 4d action, we need to expand the relevant fields in eigenbasis of the Laplacian, and then perform dimensional reduction. More precisely, we denote by $\psi_n$ a real 0-form basis of the Laplacian, normalised as
\begin{align}
\Delta_0 \psi_n &\equiv -c_n^2 \psi_n \\
\frac{1}{V_S} \int_S \psi_n \wedge * \psi_m &\equiv \delta_{nm}\,,
\end{align}
where $V_S$ stands for the volume of the four-cycle $S$.  
As said before, $\a$ should contain the eigenmodes of the gauge vector field $A$. Now, given the relation \eqref{generalalpha} and the fact that $[\Delta, \d^c] =0$, if the function $g$ is an eigenmode of the Laplacian so will be $\alpha$. Therefore, one naturally expands $\a$ as 
\be
\a =\frac{2}{V_S} \sum_{n\neq 0} a_n(x)\,  \d^c \frac{\psi_n}{c_n}\,,
\ee
where $a_n(x)$ are interpreted as canonically-normalised 4d fields, which are eventually going to acquire a vev. Additionally, we can interpret the function $\varphi$ defined in \eqref{varphi} in terms of the internal profile of the Higgs-field zero mode. More precisely, near the wall of stability we have that
\be
\varphi = |\phi(x)|^2 \frac{1}{V_S} \sum_n m_n \psi_n\,,
\label{ap:varphi}
\ee
where $m_n \in \IR$ and $\phi(x)$ is the 4d charged field whose vev generates a T-brane profile of the form (\ref{NilpPhi}). On the one hand, the fact that $\phi$ is canonically normalised translates into $m_0 =1$. On the other hand, the fact that we obtain a finite quartic coupling for this field when we plug \eqref{ap:varphi} into \eqref{Daction} translates to the fact that the sum $\sum_n m_n^2$ must converge. Finally, one may easily extend this decomposition to a more general non-nilpotent-Higgs-field profile. Here for simplicity we will focus on the nilpotent case.

Plugging both expansions in the above action we obtain 
\begin{align}
S &\supset \frac{1}{2V_S}  \int_{\mathbb{R}^{1,3}} {\rm{d}}^4x \; \bigg( \big( c V_S + |\phi|^2
\big)^2 + \sum_{n \neq 0} \left(4 c_n a_n - m_n |\phi|^2   \right)^2 \bigg) \label{4daction}\,,
\end{align}
which is nothing but eq. \eqref{DiffD} expanded in a basis of eigenmodes of the Laplacian. In other words, we have that at the wall there are cubic couplings of the form $a_n|\phi|^2$.  If now $c\neq 0$ and $\phi$ develops a vev to cancel the first term, that is the usual 4d D-term, the Kaluza-Klein modes of the gauge vector field must also do so. In particular we have that 
\begin{align}
<a_n> & = \frac{m_n}{4c_n} |\phi|^2\,.
\end{align}
As the $m_n$ are bounded from above, these vev's for the KK modes will typically decrease as their mass $c_n$ increases.

\section{Examples of wall crossing for coincident branes}\label{ap:examples}

As a proof of existence, we will construct different examples of 4-cycles inside a compact Calabi-Yau showing the properties discussed in section \ref{coincbranes}. Consider the toric ambient space $\mathbb{P}_1 \times \mathbb{P}_1 \times \mathbb{P}_2$, where we label coordinates and divisor classes as given in table \ref{P1P1P2}.
\begin{table}[h]
	\centering
	\begin{tabular}{ccccccc}
		$x_1$ & $x_2$ & $x_3$ & $x_4$ & $x_5$ & $x_6$ & $x_7$ \\\hline
		1 & 1 & 0 & 0 & 0 &0 & 0 \\\hline
		0 & 0 & 1 & 1 & 0 & 0 & 0 \\\hline
		0 & 0 & 0 & 0 & 1 & 1 & 1 \\\hline
		$\uparrow$ & & $\uparrow$ & & $\uparrow$\\
		$H_1$ & & $H_2$ & & $H_3$\\
	\end{tabular}
	\caption{Ambient space $\mb{P}_1 \times \mb{P}_1 \times \mb{P}_2$.}
	\label{P1P1P2}
\end{table}
Using the Stanley-Reisner ideal, we can read off that the only non-vanishing intersection product in the ambient space is given by $H_1 \cdot H_2 \cdot H_3^2 = 1$. We define a Calabi-Yau 3-fold $X$ inside this ambient space by the zero locus of the most general polynomial in the class $[X] = 2 H_1 + 2 H_2 + 3 H_3$. One may check that $X$ is non-singular. Using Lefshetz hyperplane theorem we know that $H^{1,1} (\mb{P}_1 \times \mb{P}_1 \times \mb{P}_2) \cong H^{1,1}(X)$, such that $X$ inherits the K\"ahler form
\begin{align}
J = v_1 \, H_1 + v_2 \, H_2 + v_3 \, H_3\, , \qquad \qquad v_i \geq 0
\end{align}
from the ambient space. Similarly, we have $H^{0,1}(X) = H^{0,1} (\mb{P}_1 \times \mb{P}_1 \times \mb{P}_2) = 0$. In the following we will show different wall-crossing phenomena present on three 4-cycles inside the Calabi-Yau.

\subsection*{Decay}
First, consider the 4-cycle $S$ defined by the vanishing locus $S= \{x_5 + x_6 + x_7 = 0\}$. Using the adjunction formula, we compute its total Chern class as
\begin{align}
c(S) &= \frac{c(X)}{[S]} = \frac{c(\mb{P}_1 \times \mb{P}_1 \times \mb{P}_2)}{[X] \, [S]} \\
&= 1 - H_3 + \cdots \nonumber,
\end{align}
from which we can read off in particular that $S$ is negatively curved, $\mc{R}  = - c_1(K_S)= c_1(S) = -H_3$. In the notation of section \ref{coincbranes}, we take
\begin{align}
\mc{M} &= H_1 \\
\Rightarrow \mc{P} &= \mc{M}^{-1} \otimes \mc{K_S}^2 = 2 H_3 - H_1\,,
\end{align}
where we can identify line bundles and their Chern classes, because $h^{0,1}=0$ and therefore $\mr{Pic} (S) \cong H^{1,1}(S) \cap H^2 (S, \mathbb{Z})$. To determine the physical spectrum of the coincident branes we need to compute the zeroth and first cohomologies of $\mc{M}$ and $\mc{P}$. We can simply read off the zeroth cohomologies from the toric data, where wee see, in particular, that $\mc{M}$ is effective whereas $\mc{P}$ is not. To determine the first cohomology groups we use \emph{cohomCalg} \cite{cohomCalg:Implementation,Blumenhagen:2010pv}, and in summary we have
\begin{align}
h^\bullet (\mc{M}) &= \big(2, 0, 0 \big) \\
h^\bullet (\mc{P}) &= \big(0, 0, 0 \big)\,.
\end{align}
From here we see that T-branes can only be stable on one side of the wall. Moreover, from 
\begin{align}
\xi &= -2 \int_S c_1(\mc{L}) \w J = - \frac{1}{2} \int_S \Big( c_1(\mc{M}) - c_1(\mc{P}) \Big) \w J \nonumber\\
&= 2 v_1 - v_2 - 2 v_3\,,
\end{align}
we see that the Fayet-Ilioupoulos term can indeed acquire both signs depending on the position in K\"ahler moduli space. Notice that $\int_S c_1^2(K_S) = 0$ and $I = 2$, in agreement with the necessary condition of section \ref{coincbranes} for a decay.

\subsection*{T-brane to T-brane crossing}
Let us repeat the analysis of the last subsection for the different combination of 4-cycle $S$ and line bundle $\mc{M}$ given by
\begin{align}
[S] &= 2 H_1 + 3 H_3 \\
\mc{M} &= H_1 + 4 H_3 \\
\Rightarrow \mc{P} &= 3 H_1 + 2 H_3\,,
\end{align}
where $S$ should be defined for instance by the most general polynomial in the given class in order to be non-singular. The line bundle cohomologies are given by
\begin{align}
h^\bullet (\mc{M}) &= \big(30, 0, 0 \big) \\
h^\bullet (\mc{P}) &= \big(24, 0, 0 \big)
\end{align}
and the Fayet-Ilioupoulos
\begin{align}
\xi &= -6 v_1 - 3 v_2 + 2 v_3\,.
\end{align}
From the above we read off that the Fayet-Ilioupoulos term can acquire both signs, and T-branes are stable on both sides, due to the condensation of either the modes of $\mc{M}$ or of $\mc{P}$.

\subsection* {T-brane to T-brane or bound state of gauge field}
Last, consider
\begin{align}
[S] &= 2 H_1 + 2 H_3 \\
\mc{M} &= 3 H_3 \\
\Rightarrow \mc{P} &= 2 H_1 + H_3\,,
\end{align}
where the bundle cohomologies are given by
\begin{align}
h^\bullet (\mc{M}) &= \big(10,1, 0 \big) \\
h^\bullet (\mc{P}) &= \big(9, 0, 0 \big)\,,
\end{align}
and the Fayet-Ilioupoulos is given by
\begin{align}
\xi &= -4 v_1 - v_2 + 2 v_3\,,
\end{align}
which can acquire both signs depending on the position in K\"ahler moduli space. We read off that on one side of the wall T-branes are stable, whereas at the other side we may either have T-brane bound states, non-Abelian gauge profiles or a combination of the two.

\end{document}